\documentclass[lettersize,journal]{IEEEtran}
\IEEEsettextwidth{0.6667in}{0.6667in}
\IEEEsetsidemargin{c}{0pt}
\usepackage[utf8]{inputenc}
\usepackage[T1]{fontenc}
\usepackage{amsmath,amsfonts,amssymb}
\usepackage{array}
\usepackage{textcomp}
\usepackage{stfloats}
\usepackage{url}
\usepackage{graphicx}
\usepackage[colorlinks=true,linkcolor=blue,citecolor=blue,urlcolor=blue]{hyperref}
\makeatletter
\def\@IEEENORMtitlevspace{-0.375\baselineskip}
\def\@IEEEMINtitlevspace{-0.5\baselineskip}
\makeatother

\begin{document}

\title{Ultra-Short-Term Photovoltaic Ramp Forecasting via Physics-Guided Generative Sky Video Prediction}

\author{Siyuan~Wang,~\IEEEmembership{Member,~IEEE,}
        Fengqi~You,~\IEEEmembership{Senior~Member,~IEEE}
\thanks{Corresponding author: Fengqi You.}
\thanks{S. Wang and F. You are with Duffield College of Engineering, Cornell University AI for Science Institute and Cornell AI for Sustainability Initiative (CAISI), Cornell University, Ithaca, New York 14853, USA. (E-mail: siyuan.wang@cornell.edu, fengqi.you@cornell.edu)}}

\maketitle

\begin{abstract}
Minute-scale photovoltaic (PV) ramps caused by rapidly evolving cloud occlusion pose challenges to short-timescale power balancing and require reliable advance warning. This paper develops an event-oriented ultra-short-term PV forecasting framework that predicts future cloud evolution as an intermediate representation linking sky observations to PV output. PhyDiffNet combines physics-informed cloud-motion modeling with generative video prediction, while RaPVFormer integrates observed and predicted sky states, sun-location information, and historical PV measurements to forecast PV trajectories and ramp events up to 16 minutes ahead. Evaluation on the SKIPP'D dataset shows that the proposed framework achieves a Critical Success Index of 96.2\% for ramps beginning 1--4 minutes ahead and retains 84.9\% for those beginning 13--16 minutes ahead, outperforming all deployable baselines across the evaluated lead-time intervals. At a validation-calibrated 95\% reliability target, it reduces the required fast upward reserve margin by 34.2\% relative to the Persistence Model and by 11.6\% relative to the strongest deployable baseline under the adopted reserve metric. These results show that explicit future-cloud prediction improves advance warning of PV ramps and translates forecasting gains into lower short-timescale balancing margins under a common reliability target.
\end{abstract}


\section{Introduction}

\subsection{Motivation}
\IEEEPARstart{P}{hotovoltaic} (PV) generation is becoming an increasingly important component of low-carbon power systems, but its output can change rapidly when moving clouds obscure or reveal the sun \cite{IEA_PVPS_Snapshot2025,zangImprovingUltrashorttermPhotovoltaic2024}. High-resolution measurements document substantial solar-irradiance ramps at one-minute and sub-minute timescales \cite{JainEtAl_PVVariability2020}, and such fast irradiance transitions can translate into abrupt PV power ramps \cite{hendrikxAllSkyImagingbased2024}. At high PV penetration, these minute-scale ramps complicate balancing and dispatch because their timing and magnitude may not be captured by forecasts designed for longer horizons or smoother weather evolution \cite{wenDeepLearningBased2021,chengMitigatingImpactPhotovoltaic2024}. More accurate advance warning of these events can support reserve activation, storage scheduling, ramp-rate control, and other short-timescale operational actions \cite{fregosiAnalysisStorageRequirements2023}. This motivates forecasting methods that explicitly target the occurrence and evolution of PV ramp events rather than only minimizing average pointwise power-prediction error.

The main source of uncertainty at minute-scale horizons is the local evolution of clouds around the solar disk. Numerical weather prediction (NWP) and satellite observations provide valuable information for longer horizons and larger spatial domains, but their spatial and temporal resolutions are often insufficient to resolve the cloud-edge motion and solar occlusion that drive very short-term PV ramps \cite{chuIntrahourIrradianceForecasting2021,liuDayaheadPhotovoltaicPower2025,chuReviewDistributedSolar2024}. Ground-based full-sky cameras directly observe these localized cloud dynamics and therefore provide a complementary sensing modality for intra-hour solar forecasting \cite{palettaAdvancesSolarForecasting2023,chenForecastingBasedPowerRampRate2019}. The forecasting problem, however, is not only to interpret the current sky image, but also to anticipate how cloud morphology and its position relative to the sun will evolve over the next several minutes.

Recent advances in physics-informed learning, generative video prediction, and attention-based sequence modeling provide a pathway to represent these future sky states. For PV ramp forecasting, the value of such models depends less on photorealistic video generation itself than on whether the predicted frames preserve the cloud structures, motion, and solar-occlusion patterns that determine subsequent PV output. Accordingly, this work treats future sky-video prediction as an intermediate representation for the primary task of ultra-short-term PV power and ramp-event forecasting.

\subsection{Literature Review and Research Gap}
PV forecasting methods span NWP, satellite remote sensing, local ground-based sensing, and hybrid approaches \cite{chuIntrahourIrradianceForecasting2021}. NWP is typically used at hourly or longer horizons \cite{liuDayaheadPhotovoltaicPower2025}, while satellite-based approaches commonly provide intra-day forecasts at resolutions of several minutes \cite{chuReviewDistributedSolar2024}. For ultra-short-term forecasting, full-sky imagery can capture local cloud motion and solar occlusion at substantially finer spatial and temporal scales \cite{palettaAdvancesSolarForecasting2023}. Early sky-image approaches used image segmentation \cite{westShorttermIrradianceForecasting2014,zhangNovelIntrahourPV2024} or convolutional neural networks such as SUNSET \cite{sunSolarPVOutput2018} to map observed sky conditions to irradiance or PV output. Subsequent models introduced classification-based learning \cite{martinezlopezUsingSkyclassificationImprove2024}, recurrent spatiotemporal representations such as ConvLSTM \cite{ShiEtAl_ConvLSTM_NIPS2015}, and event-oriented frameworks such as ECLIPSE \cite{palettaECLIPSEEnvisioningCLoud2022}. Recent studies have further advanced image-based ultra-short-term PV forecasting through sky-image-derived deep decomposition \cite{liuSkyImageDerivedDeep2024} and transformer-based spatiotemporal learning from observed sky-image sequences \cite{rastgooUltraShortTermSolar2025}. These studies demonstrate the use of learned spatiotemporal representations of observed sky-image sequences for ultra-short-term PV forecasting. A remaining question is whether explicitly representing future cloud evolution can improve the anticipation of rapid PV ramp events.

Physics-informed architectures such as PhyDNet \cite{LeGuenThome_CVPR_2020,LeGuenThome_CVPRW2020_IrradianceFisheye} introduce physical structure into latent cloud-motion dynamics, while generative sequence models such as VideoGPT \cite{yanVideoGPTVideoGeneration2021} and SkyGPT \cite{nieSkyGPTProbabilisticUltrashortterm2024} generate possible future visual states for solar forecasting. For ramp-oriented forecasting, the remaining challenge is to preserve cloud-edge geometry, motion, and solar-occlusion evolution over multiple minutes with sufficient fidelity to support accurate prediction of ramp occurrence, timing, direction, and magnitude.

Two issues are therefore particularly important for PV ramp events. First, fast-evolving clouds undergo translation, deformation, formation, and dissipation, so multi-minute future-sky predictions can progressively lose precisely the cloud-edge and occlusion information that governs abrupt changes in PV output. Second, point-forecasting accuracy alone does not characterize whether a model can provide reliable advance warning of an operationally relevant ramp. Ramp-oriented evaluation additionally requires event-level measures of detection success, start and end times, and ramp magnitude, together with an assessment of how these quantities degrade as event lead time increases. In this work, future-sky prediction is consequently treated as an intermediate representation whose value is assessed primarily through downstream PV and ramp quantities, while conventional image- and video-quality metrics are retained as supporting diagnostics.

\subsection{Contributions}
To address these challenges, we develop a two-stage ramp-oriented forecasting framework in which predicted future sky states bridge visual observations and downstream PV-ramp forecasts. The main contributions are as follows:

(1) We formulate explicit future-cloud prediction as an intermediate representation for event-oriented ultra-short-term PV forecasting. Unlike approaches that map observed sky-image sequences directly to future power, the proposed formulation uses predicted future sky states to anticipate the occurrence, timing, direction, and magnitude of minute-scale PV ramps.

(2) We develop an integrated PhyDiffNet--RaPVFormer architecture for this formulation. PhyDiffNet combines physics-informed cloud-motion modeling with video-conditional diffusion to preserve irradiance-relevant cloud motion and cloud-edge evolution, while RaPVFormer fuses observed and predicted sky states, deterministic sun-location information, historical PV output, and ramp-sensitive learning to forecast future PV trajectories.

(3) We characterize PV-ramp forecasting capability as a function of true ramp-onset lead time and quantify its short-timescale operational implication. Ramp detection, start/end timing, and magnitude are evaluated over 1--4, 5--8, 9--12, and 13--16 minute onset-lead intervals, while a validation-calibrated reliability--reserve analysis links residual PV forecast errors to fast upward balancing margins. This evaluation connects forecasting accuracy directly with advance ramp-warning capability and balancing requirements under common reliability targets.

\section{PV Ramp Forecasting Framework}

\subsection{Overview of the Framework}
\begin{figure}[!t]
\centering
\includegraphics[width=\columnwidth]{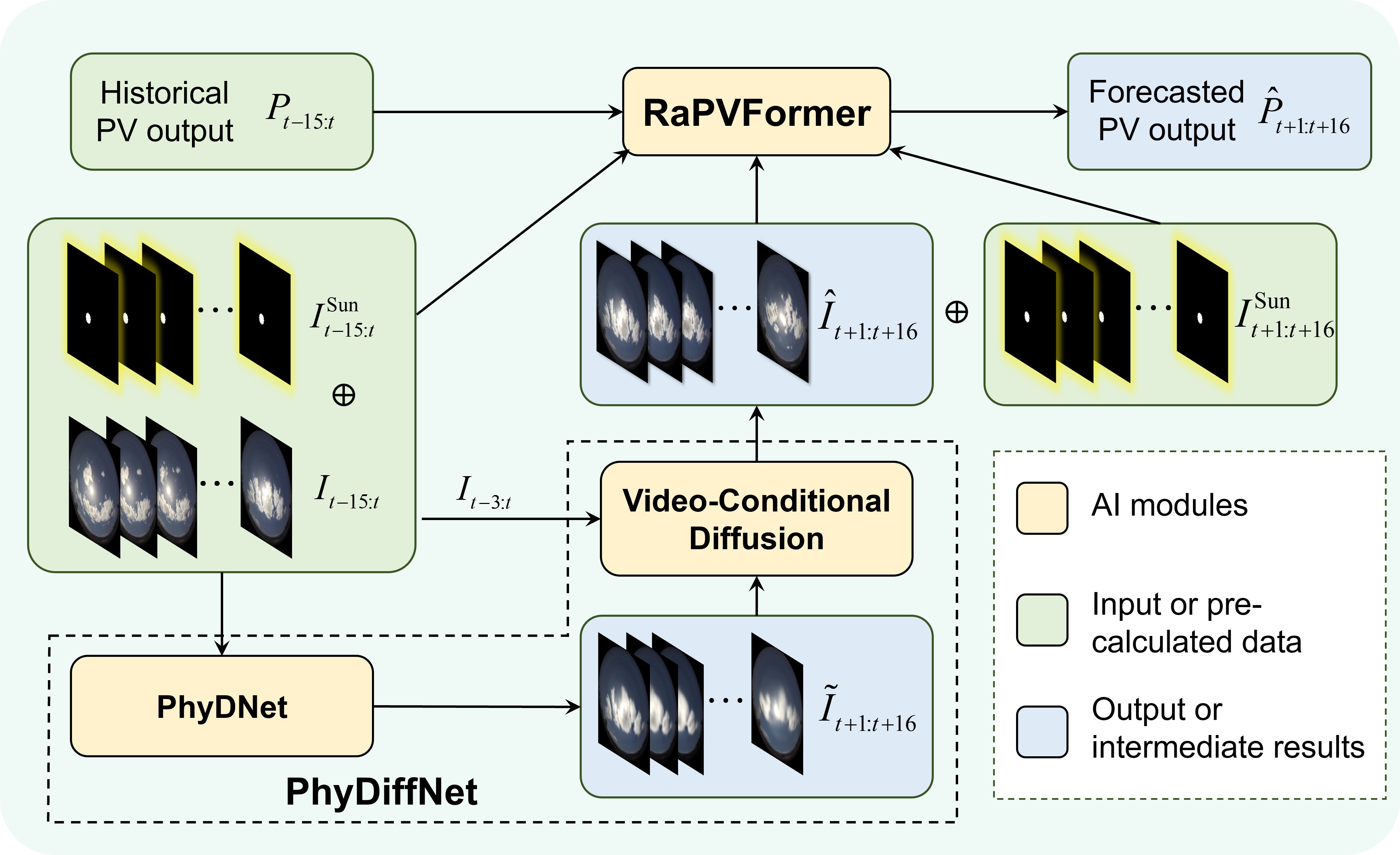}
\caption{Data flow of the AI framework for full-sky frames prediction and PV output forecasting.}
\label{fig:framework_overview}
\end{figure}

The proposed framework consists of two main components: PhyDiffNet for high-fidelity, motion-sensitive sky video prediction, including PhyDNet and video-conditional diffusion modules; and RaPVFormer for ramp-aware PV output forecasting. The data flow of the whole AI framework is shown in Fig.~\ref{fig:framework_overview}. Detailed code implementation is available in the project repository \cite{WangYou_PVRamp_2026}.

The historical video frames from the past 16 minutes at 1-minute resolution, denoted as $I_{t-15:t}$, together with the corresponding precomputed sun-mask channel $I^{Sun}_{t-15:t}$, are used to initialize PhyDNet, which models cloud dynamics in a semantic latent space. During autoregressive decoding, the precomputed future sun masks $I^{Sun}_{t+1:t+16}$ are supplied at the corresponding forecast steps. This process generates a coarse prediction of blurred full-sky frames for the next 16 minutes, denoted as $\tilde I_{t+1:t+16}$.

\begin{equation}
\tilde I_{t+1:t+16} = \mathrm{PhyDNet}\left(I_{t-15:t}, I^{Sun}_{t-15:t+16}\right)
\label{eq:phydnet}
\end{equation}

The sun-mask sequence can be precomputed using the algorithm from \cite{niePVPowerOutput2020}, based on fisheye camera parameters, geographical coordinates, and time. Because solar position is deterministic for a given location and time once the camera geometry is known, $I^{Sun}_{t+1:t+16}$ is available at forecast issuance and can be used during inference without requiring future sky observations.

Next, the historical video frames from the past 4 minutes $I_{t-3:t}$ and the predicted blurred full-sky frames $\tilde I_{t+1:t+16}$ are used as conditional inputs to refine the cloud textures in the 16-minute future predictions, resulting in high-fidelity frames denoted as $\hat I_{t+1:t+16}$.

\begin{equation}
\hat I_{t+1:t+16} = \mathrm{Diffusion}\left(I_{t-3:t}, \tilde I_{t+1:t+16}\right)
\label{eq:diffusion}
\end{equation}

Finally, the historical video frames $I_{t-15:t}$, corresponding historical PV outputs $P_{t-15:t}$, the refined future frames $\hat I_{t+1:t+16}$, and the associated sun-mask sequence $I^{Sun}_{t-15:t+16}$ are input into RaPVFormer to predict the future PV outputs $\hat P_{t+1:t+16}$ using irradiance changes captured in the predicted sky frames.

\begin{multline}
\hat P_{t+1:t+16} = \mathrm{RaPVFormer}\big(I_{t-15:t}, \hat I_{t+1:t+16}, \\ I^{Sun}_{t-15:t+16}, P_{t-15:t}\big)
\label{eq:rapvformer}
\end{multline}

\subsection{PhyDiffNet for High-Fidelity, Motion-Sensitive Sky Video Prediction}
Following the PhyDNet formulation, cloud dynamics are modeled in a semantic latent space, where physical and residual components are represented separately.

\begin{figure}[!t]
\centering
\includegraphics[width=\columnwidth]{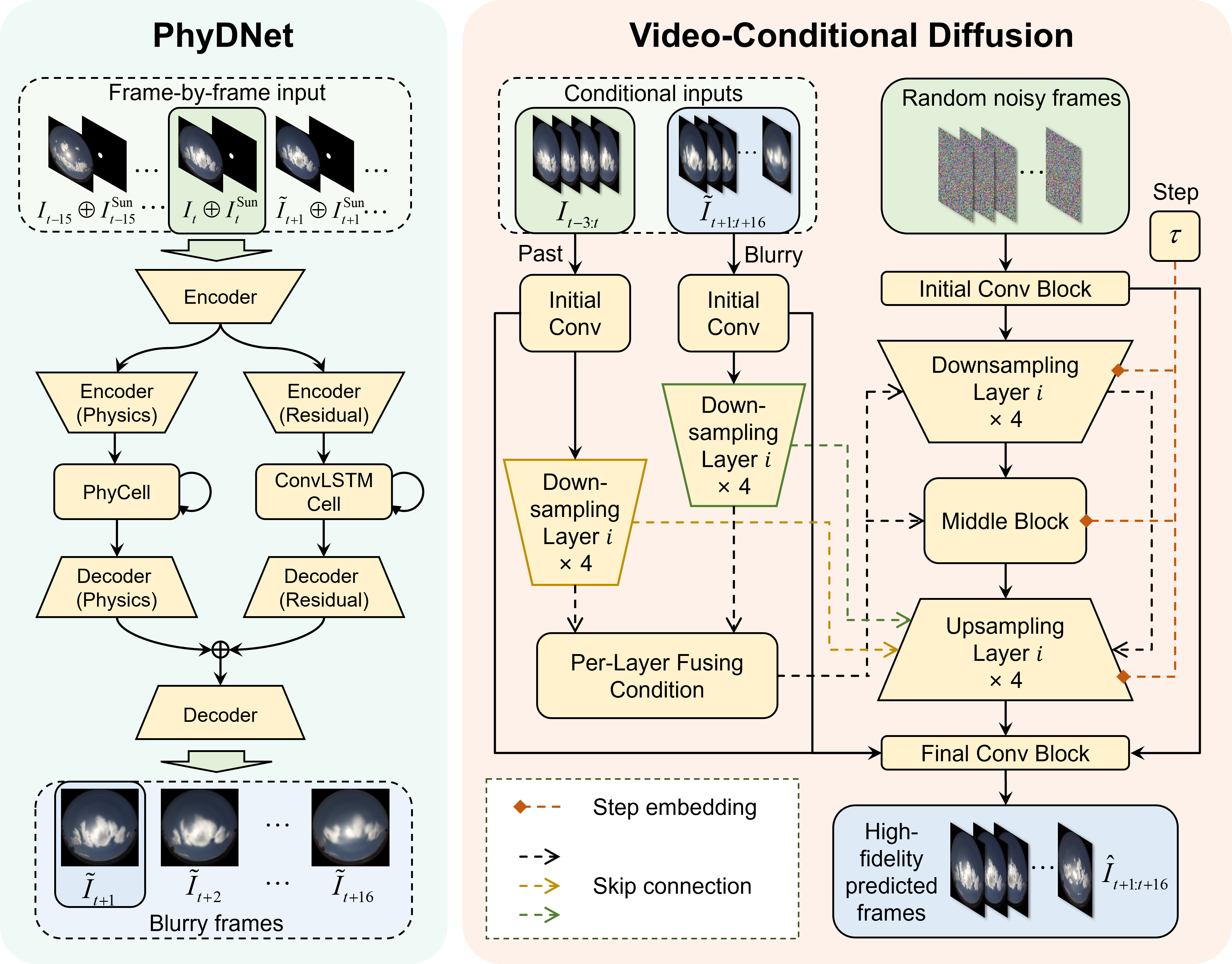}
\caption{PhyDiffNet framework integrating PhyDNet with a video-conditional diffusion module for high-fidelity, motion-aware full-sky video frame prediction.}
\label{fig:phydiffnet_architecture}
\end{figure}

The partial differential equation (PDE) that governs the dynamics in the latent space $h$ can be decomposed into two components: the physical dynamics $h^{Phy}$ and the residual dynamics $h^{Res}$, as follows:

\begin{equation}
\frac{\partial h}{\partial t} = \frac{\partial h^{Phy}}{\partial t} + \frac{\partial h^{Res}}{\partial t}
\label{eq:pde_decomp}
\end{equation}

\noindent Here, $h^{Phy}$ captures the primary motion governed by physical constraints, while $h^{Res}$ accounts for components not modeled by the physical system. To implement this idea, PhyDNet introduces a dual-branch architecture to model the latent dynamics of video frames \cite{LeGuenThome_CVPR_2020,LeGuenThome_CVPRW2020_IrradianceFisheye} as shown in Fig.~\ref{fig:phydiffnet_architecture}. One branch incorporates a physics-guided recurrent cell (PhyCell), specifically designed to model cloud motion using PDEs.

\begin{equation}
h_{t+1} = (1-K_t)\odot \tilde h_{t+1} + K_t\odot \mathrm{Enc}\left(I_t\oplus I_t^{Sun}\right)
\label{eq:phycell_combine}
\end{equation}

\begin{equation}
\tilde h_{t+1} := h_t + \Phi(h_t)
\label{eq:phycell_phys}
\end{equation}

\noindent where $\tilde h_{t+1}$ denotes the predicted latent state at the next time step $t+1$, computed using the physical model only; $\mathrm{Enc}(\cdot)$ denotes the encoded representation of the current input frame and sun-mask channel; $\odot$ denotes the Hadamard product; $\oplus$ denotes the channel concatenation; $\Phi(h_t)$ is the physical predictor term, approximating spatial dynamics using learned differential operators and modeled via convolutions as \eqref{eq:phy_conv}; $K_t$ is the gating term controlling the balance between physics prediction and observation correction and is approximated by a neural network as \eqref{eq:gate_conv}.

\begin{equation}
\Phi(h_t) = \sum_{i+j\le q} c_{i,j}\frac{\partial^{i+j}h_t}{\partial x^i \partial y^j} \approx \sum_{i+j\le q} c_{i,j} * \nabla_{i,j} h_t
\label{eq:phy_conv}
\end{equation}

\begin{equation}
K_t = \sigma\left(W_h * h_t + W_u * \mathrm{Enc}\left(I_t\oplus I_t^{Sun}\right) + b\right)
\label{eq:gate_conv}
\end{equation}

\noindent where $c_{i,j}$, $W_h$, $W_u$, $b$ are learnable coefficients; $\nabla_{i,j}$ is the convolutional kernel that approximates the differential operator; and the sigmoid function $\sigma(\cdot)$ constrains $K_t$ to $[0,1]$. The gate is computed from the current input encoding and the previous hidden state $h_t$. The second branch consists of a standard ConvLSTM, which learns the residual motion patterns not captured by the physics-based model. Finally, the output features from both branches are summed and decoded to generate the predicted future frame.

In the PhyDNet module, historical video frames $I_{t-15:t}$ and corresponding sun-masks $I^{Sun}_{t-15:t}$ are sequentially input into the model, which subsequently generates predicted frames via its decoder.

Furthermore, the loss function incorporated the Structural Similarity Index Measure (SSIM) \cite{wang2004ssim} to emphasize the preservation of image structure and texture.

\begin{equation}
\mathcal{L}_{frame}(I_t,\tilde I_t) = \alpha_F \mathcal{L}_{SSIM}(I_t,\tilde I_t) + (1-\alpha_F)\mathcal{L}_1(I_t,\tilde I_t)
\label{eq:loss_frame}
\end{equation}

\begin{equation}
\mathcal{L}_{SSIM}(I_t,\tilde I_t) = 1 - \mathrm{SSIM}(I_t,\tilde I_t)
\label{eq:loss_ssim}
\end{equation}

\noindent where $\mathrm{SSIM}(\cdot)$ denotes the structural similarity index measure, defined in the supplementary file \cite{WangYou_PVRamp_2026}; loss function $\mathcal{L}_{SSIM}$ is used to measure image structural similarity and $\mathcal{L}_1$ is used to reflect pixel-level differences; $\alpha_F$ balances these two terms. Training additionally uses differential-moment regularization, whose  weight is provided in supplementary file \cite{WangYou_PVRamp_2026}.

PhyDiffNet uses video-conditional diffusion to refine coarse PhyDNet outputs and recover finer cloud structures and textures. The module is adapted from denoising diffusion probabilistic models (DDPMs) \cite{ho2020ddpm}. DDPMs define a Markov forward process that gradually adds Gaussian noise to the clean data sample $x_0$ over $\Gamma$ steps:

\begin{multline}
q(x_\tau|x_{\tau-1}) = \mathcal{N}\left(x_\tau;\sqrt{1-\beta_\tau}\,x_{\tau-1},\,\beta_\tau I\right), \\ \tau=1,2,\cdots,\Gamma
\label{eq:forward_diffusion}
\end{multline}

\noindent where $x_\tau$ denotes the noisy sample at diffusion step $\tau$ and $\beta_\tau \in (0,1)$ is the noise variance. We use $\Gamma=1000$ diffusion steps with the cosine noise schedule proposed by Nichol and Dhariwal \cite{nichol2021improved}, setting the offset parameter to $s=0.008$.

The video-conditional diffusion module refines blurry frames $\tilde I_{t+1:t+16}$ using a diffusion-based denoising process. Specifically, the predicted blurry frames $\tilde I_{t+1:t+16}$ from PhyDNet and the last observed four frames $I_{t-3:t}$ are used as the conditional input $x_{cond} := (I_{t-3:t}, \tilde I_{t+1:t+16})$. At each diffusion step, the denoiser model predicts the injected noise, denoted as $\epsilon_\theta(x_\tau,\tau,x_{cond})$, based on both the noisy sample $x_\tau$ and the conditional input $x_{cond}$:

\begin{equation}
\mathcal{L}_{DDPM}=\mathbb{E}_{x_0,\epsilon,\tau}\left[\left\|\epsilon-\epsilon_\theta(x_\tau,\tau,x_{cond})\right\|^2\right]
\label{eq:ddpm_loss}
\end{equation}

\begin{figure}[!t]
\centering
\includegraphics[width=\columnwidth]{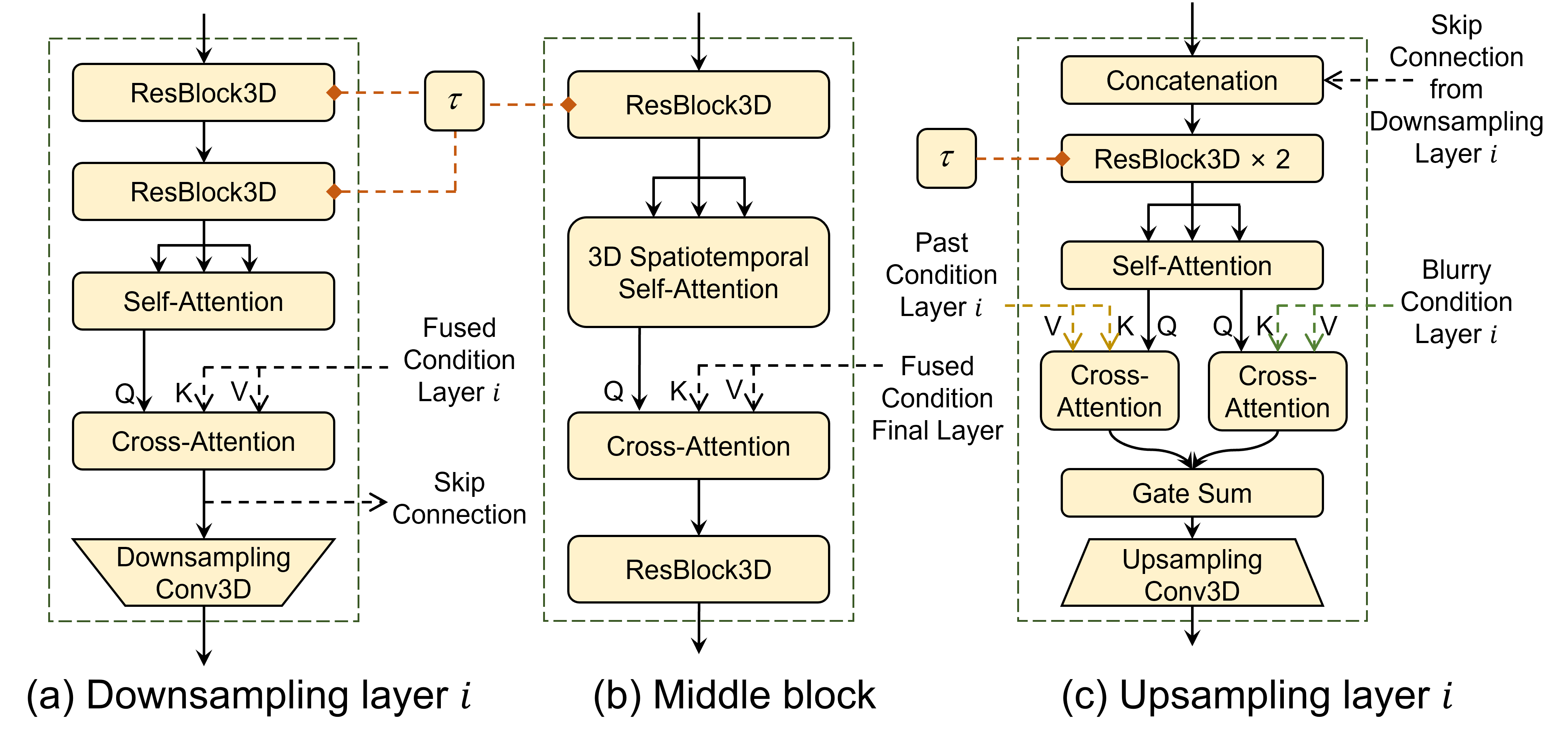}
\caption{Detailed design of blocks in the backbone branch of the 3D U-Net, including (a) downsampling layer; (b) middle block; and (c) upsampling layer. The downsampling layers are numbered sequentially, while the upsampling layers are numbered in reverse order.}
\label{fig:unet_block_design}
\end{figure}

The core denoising network $\epsilon_\theta$ is a spatiotemporal 3D U-Net that jointly models temporal and spatial dependencies. The architecture comprises a backbone branch and a conditional-input branch. The backbone uses a U-shaped encoder--decoder structure with symmetric downsampling and upsampling paths. The conditional-input branch processes past and blurry future frames independently and concatenates their features at multiple resolutions. Cross-attention injects these fused features into the backbone, integrating temporal cues and coarse future structures. Fig.~\ref{fig:unet_block_design} shows the blocks used in the backbone branch.

In the downsampling path, the model progressively reduces the spatial resolution while preserving the temporal dimension, with the layer structure shown in Fig.~\ref{fig:unet_block_design}(a). Each resolution level comprises two residual 3D convolutional blocks, each consisting of 3D convolutions, nonlinear activation function, and in-block skip connection, following the ResNet architecture \cite{heDeepResidualLearning2015}. Additionally, a linear self-attention layer and a linear cross-attention layer are applied at each resolution level, incorporating the fused conditional features with matching spatial dimensions. Compared to the standard attention realization in \eqref{eq:std_attn}, the linear attention realization \eqref{eq:lin_attn} reduces the computational and memory complexity from quadratic to linear in sequence length.

\begin{equation}
\mathrm{Attn}(Q,K,V)=\mathrm{softmax}\left(\frac{QK^\top}{\sqrt{d}}\right)V
\label{eq:std_attn}
\end{equation}

\begin{equation}
\mathrm{LinAttn}(Q,K,V)=\frac{\phi(Q)\left(\phi(K)^\top V\right)}{\phi(Q)\left(\phi(K)^\top \mathbf{1}\right)}
\label{eq:lin_attn}
\end{equation}

\noindent where $Q$, $K$ and $V$ denote queries, keys and values, respectively, and $d$ is the key dimensionality. Standard attention \eqref{eq:std_attn} constructs an $O(L^2)$ attention matrix, where $L$ is the number of tokens. Linear attention \eqref{eq:lin_attn} applies kernel feature maps to queries and keys, avoiding the full $QK^\top$ matrix \cite{katharopoulos2020transformers}. We use the positive feature map $\phi(x) := \mathrm{ELU}(x)+1$ for both queries and keys. The main attention computation scales as $O(LDd_v)$, where $D$ and $d_v$ denote the feature-map and value dimensions, respectively. For fixed feature dimensions, its cost grows linearly with the number of tokens.

The downsampling layers of the conditional-input branch are similar to those of the backbone branch, as shown in Fig.~\ref{fig:unet_block_design}(a), but exclude step embeddings as well as self- and cross-attention blocks.

The middle block sits at the lowest resolution and integrates full spatiotemporal context. It applies a linear self-attention block and a linear cross-attention block, both defined over 3D feature maps, as shown in Fig.~\ref{fig:unet_block_design}(b).

The upsampling path mirrors the downsampling path structure, with the layer structure shown in Fig.~\ref{fig:unet_block_design}(c). It restores spatial resolution and incorporates skip connections from downsampling layers at the corresponding resolution level. Each upsampling layer applies self-attention, and further incorporates gated cross-attention mechanisms to inject conditional information. Formally, at each level $i$, the update is:

\begin{multline}
h^{(i)}=u^{(i)}+\sigma\!\left(g^{(i)}_{past}\right)\cdot \mathrm{Attn}^{(i)}_{past}\!\left(u^{(i)},F^{(i)}_{past}\right) \\ + \sigma\!\left(g^{(i)}_{blurry}\right)\cdot\mathrm{Attn}^{(i)}_{blurry}\!\left(u^{(i)},F^{(i)}_{blurry}\right)
\label{eq:gated_cross_attn}
\end{multline}

\noindent where $\sigma(\cdot)$ denotes the Sigmoid activation; $g^{(i)}_{past}$ and $g^{(i)}_{blurry}$ are learnable scalar gates modulating the influence of the past and blurry conditional-input branches, respectively. $F^{(i)}_{past}$ and $F^{(i)}_{blurry}$ are encoded features from them, respectively. $u^{(i)}$ and $h^{(i)}$ denote the input and output of the cross-attention block, respectively, and each $\mathrm{Attn}(\cdot)$ term denotes an attention increment without an internal residual connection.

\subsection{RaPVFormer for Ramp-Aware PV Output Forecasting}
\begin{figure}[!t]
\centering
\includegraphics[width=\columnwidth]{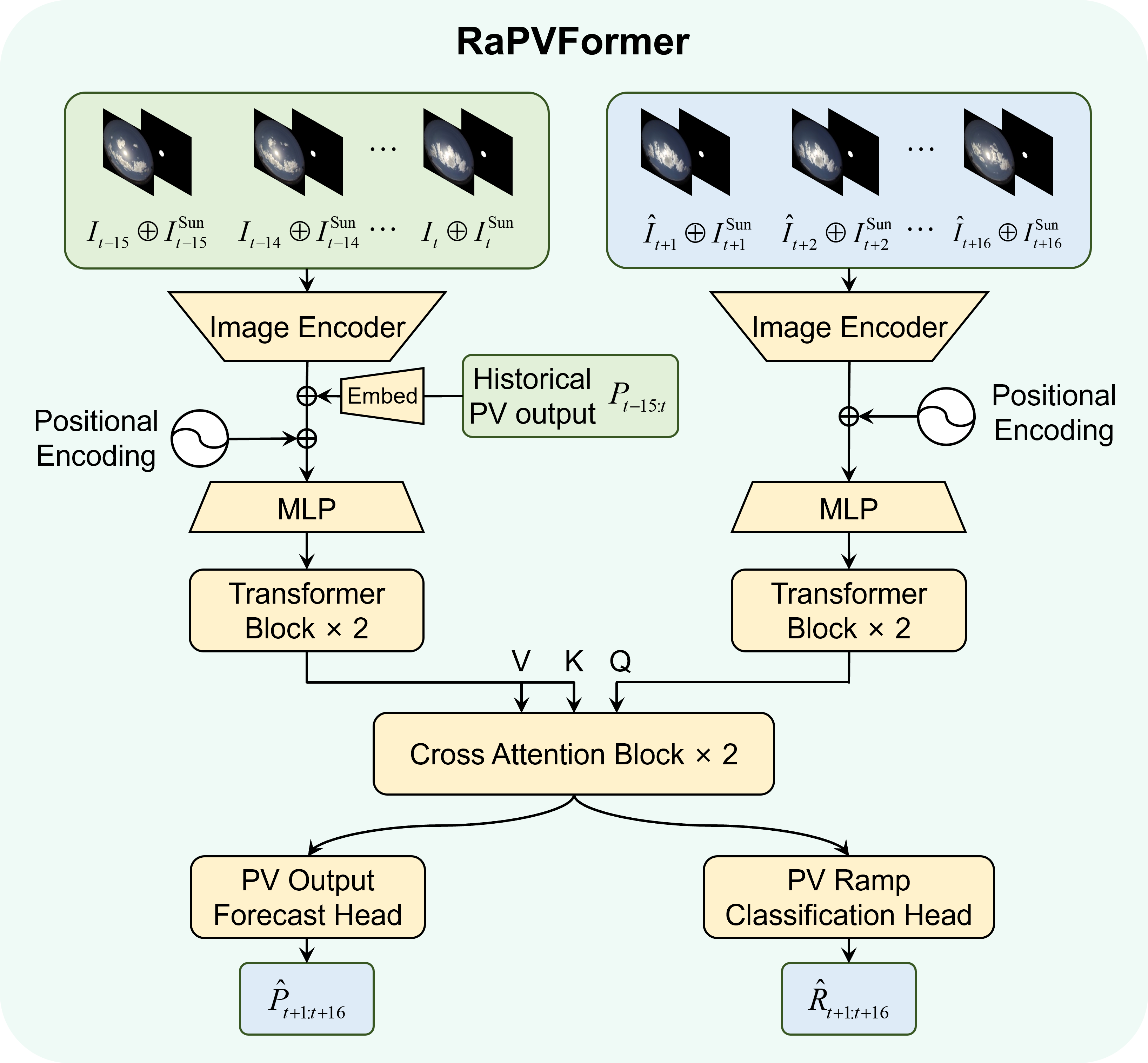}
\caption{Framework of the RaPVFormer for ramp-aware PV output forecasting.}
\label{fig:rapvformer_architecture}
\end{figure}

RaPVFormer is a transformer-based architecture designed for short-term PV forecasting under rapidly changing cloud conditions. The architecture of RaPVFormer is shown in Fig.~\ref{fig:rapvformer_architecture}. The model encodes historical and predicted sky frames ($I_{t-15:t}$ and $\hat I_{t+1:t+16}$), along with corresponding sun-masks ($I^{Sun}_{t-15:t}$ and $I^{Sun}_{t+1:t+16}$), using separate four-layer historical and future convolutional image encoders trained from scratch. Each frame is compressed into a 128-dimensional feature vector. Historical PV power values are embedded by a multi-layer perceptron (MLP) embedding and concatenated with the corresponding historical frame features, enabling the model to interpret sky appearance in the context of concurrent power output. All visual and PV features are then projected into a 256-dimensional token space and enriched with sinusoidal positional encodings to preserve temporal ordering across the input sequences.

Both historical and future token sequences are processed by separate stacks of two transformer layers, each consisting of multi-head self-attention (four heads, 256-dimensional model size) and a 512-dimensional feed-forward network. A central component of RaPVFormer is its cross-attention fusion mechanism, implemented as a stack of two cross-attention layers, each consisting of multi-head cross-attention (four heads, 256-dimensional model size) and a 1024-dimensional feed-forward network. The cross-attention mechanism fuses historical and future token representations before decoding the PV trajectory.

The fused representation is decoded by two prediction heads: one for PV-output regression $\hat P_{t+1:t+16}$ and one for three-class ramp classification $\hat R_{t+1:t+16}$ with up, down, and stable classes. The ramp-classification head serves as an auxiliary training task; its labels are derived from consecutive PV increments, including the boundary transition from $P_t$ to $P_{t+1}$. The reported event-level ramp metrics are derived from the predicted PV trajectory rather than directly from its classification logits. Training employs a composite objective combining MSE, slope-aware loss, and cross-entropy with detached focal weights to promote accurate, temporally consistent, and ramp-aware predictions, as defined below:

\begin{equation}
\mathcal{L}_{PV}=\omega_P\cdot\mathcal{L}_{MSE}+\omega_S\cdot\mathcal{L}_{Slope}+\omega_R\cdot\mathcal{L}_{Focal}
\label{eq:loss_pv}
\end{equation}

\noindent where the parameters $\omega_P$, $\omega_S$, and $\omega_R$ balance these three loss terms. The mean squared error term penalizes PV-output forecast error:

\begin{equation}
\mathcal{L}_{MSE}=\frac{1}{T}\sum_{t=1}^{T}\left(\hat P_t-P_t\right)^2
\label{eq:loss_mse}
\end{equation}

The slope loss uses an adaptive weighting scheme that increases the penalty for large ramp events, helping the model focus on critical transitions:

\begin{equation}
\mathcal{L}_{Slope}=\frac{1}{T-1}\sum_{t=2}^{T}\omega_t\left[\left(\hat P_t-\hat P_{t-1}\right)-\left(P_t-P_{t-1}\right)\right]^2
\label{eq:loss_slope}
\end{equation}

The weight $\omega_t$ is defined as follows:

\begin{equation}
\begin{aligned}
\omega_t &:=
1+\alpha_S\cdot\sigma\left(\left|P_t-P_{t-1}\right|-m_{\Delta P}\right),\\
m_{\Delta P} &:=
\underset{2\le k\le T}{\mathrm{median}}\left|P_k-P_{k-1}\right|
\end{aligned}
\label{eq:slope_weight}
\end{equation}

\noindent where $\alpha_S$ controls the weighting amplitude, $\sigma(\cdot)$ denotes the sigmoid activation function, and $m_{\Delta P}$ is the median absolute PV increment over each target sequence. This median-centered weighting assigns larger penalties to transitions whose magnitudes exceed the typical change within the sequence.

We use cross-entropy with detached focal weights to down-weight well-classified samples, denoted by $\mathcal{L}_{Focal}$:

\begin{equation}
\begin{aligned}
\mathcal{L}_{Focal}&=-\frac{1}{T}\sum_{t=1}^{T}w_t^F\log\left(\hat p_t\right),\\
w_t^F&=\operatorname{sg}\!\left[\left(1-\max\{\hat p_t,10^{-6}\}\right)^{\gamma}\right].
\end{aligned}
\label{eq:loss_focal}
\end{equation}

\noindent where $\hat p_t$ is the softmax probability of the ground-truth class and $\gamma$ controls the emphasis on hard samples. The operator $\operatorname{sg}(\cdot)$ preserves its input but blocks gradients, so backpropagation acts only through the cross-entropy term. The probability floor of $10^{-6}$ applies only when computing the detached weight.

\section{Dataset Preparation and Model Training}

\subsection{Dataset Preparation}
This study utilizes the publicly available Sky Image and Photovoltaic Power Generation Dataset (SKIPP'D) \cite{nieSKIPPDSKyImages2023,nie2017SkyImages2022}, which comprises full-sky videos recorded on the Stanford University campus using a 360-degree fisheye camera between 2017 and 2019. The dataset links are provided in the supplementary file \cite{WangYou_PVRamp_2026}. The raw data consists of full-sky images at a resolution of $2048\times2048$ pixels, accompanied by PV output measurements from a 30.1~kW array. To facilitate model training, these images were cropped and downscaled to $128\times128$ pixels, as suggested in \cite{sunSolarPVOutput2018}, while maintaining a temporal resolution of one minute. All frames captured during nighttime periods, defined as intervals from dusk to dawn with PV output less than 20\% of the installed capacity, were excluded. Additionally, frames with identical image content or power values due to sensor or camera malfunctions were removed.

Since PV ramp events are more likely to occur under partly cloudy conditions than during fully clear or overcast skies, the dataset was filtered based on the proportion of cloud coverage within the camera's field of view to improve the model's ability to learn cloud dynamic patterns. The cloud coverage ratio is calculated based on the cloud segmentation algorithm \cite{niePVPowerOutput2020}. To structure the temporal data, we applied an overlapping sliding window of 32 consecutive minutes to generate video clips. Each clip was required to include at least 20 frames with cloudiness index values between 0.1 and 0.8 to satisfy the filtering criteria. The resulting ramp-focused subset therefore emphasizes partly cloudy, dynamically evolving conditions that are most relevant to minute-scale PV ramps. After preprocessing, the final dataset consisted of 24,807 filtered full-sky images, organized into 17,333 video clips spanning 138 days. The dataset was split by day: 10\% of the days (14 days) were randomly selected for testing, another 10\% (14 days) for validation, and the remaining 80\% (110 days) were used for training. All clips from the same calendar day remained in one subset across all model stages, preventing same-day overlap between training, validation, and test sets. This split resulted in 1,775, 1,768, and 13,790 video clips for the test, validation, and training sets, respectively. To enhance generalization, random spatial rotations and flips of the whole video clips were applied as data augmentation during training; implementation details are provided in supplementary file \cite{WangYou_PVRamp_2026}.

\subsection{Model Training}
All models were trained on an NVIDIA H100 GPU with 80~GB of memory; detailed implementation settings and hyperparameters are provided in the supplementary file \cite{WangYou_PVRamp_2026}. PhyDNet was trained for 300 epochs with a batch size of 32, an initial learning rate of $10^{-4}$, and a gradually decaying teacher-forcing ratio. The video-conditional diffusion module was trained for 100,000 iterations with a batch size of 8 and a learning rate of $2\times10^{-4}$. RaPVFormer was trained for 30 epochs with a batch size of 64 and a learning rate of $10^{-4}$. Each training clip contributed equally weighted views based on the ground-truth future frames and the corresponding PhyDiffNet-generated frames, whereas validation, testing, and deployment used only generated future frames. The average end-to-end inference time was 2.62~s per forecast issuance over 100 repeated runs, well below the 1-minute forecasting resolution.

Across 1,000 forecast instances evaluated with 10 random seeds each, the PV forecasts had a mean inter-seed standard deviation of 0.43~kW, with a standard deviation of 0.59~kW across instances, indicating limited sensitivity to diffusion randomness.

\section{Evaluation Metrics}

\subsection{Primary Metrics for PV Output and Ramp Event Forecasting}
RMSE and Forecast Skill (FS) quantify continuous PV-output accuracy, while Critical Success Index (CSI), Mean Start Time Error (MSTE), Mean End Time Error (METE), and Mean Relative Magnitude Error (MRME) evaluate ramp-event detection, timing, and magnitude. The auxiliary ramp-classification head is not used directly to compute these event metrics; predicted ramp events are extracted from the forecast PV trajectory using the event definition in \eqref{eq:ramp_threshold} and then matched to observed events using \eqref{eq:event_gap}--\eqref{eq:event_matching_cost}.

RMSE is used to quantify the overall forecasting error in PV power output:

\begin{equation}
\mathrm{RMSE}:=\sqrt{\frac{1}{T}\sum_{t=1}^{T}\left(\hat P_t-P_t\right)^2}
\label{eq:rmse}
\end{equation}

The Persistence Model (PM), which assumes PV output remains constant over the forecast horizon, is used as the benchmark. Forecast Skill (FS) quantifies the improvement of the proposed method over PM:

\begin{equation}
\mathrm{FS}:=\left(1-\frac{\mathrm{RMSE}}{\mathrm{RMSE}_{PM}}\right)\times100\%
\label{eq:fs}
\end{equation}

A higher FS value indicates greater predictive accuracy relative to the baseline.

We use several event-level metrics to evaluate PV ramp predictions within the 16-minute forecast horizon. A ramp interval is defined as a continuous sequence of 1-minute PV power measurements during which the output exhibits a sustained and directionally consistent trend. That is, the incremental change $(P_{t+1}-P_t)$ preserves its sign over the interval. To prevent spurious fragmentation caused by sensor noise or short-lived fluctuations, we introduce a noise-tolerance band $\varepsilon$, set to 5\% of installed capacity, such that variations within $|P_{t+1}-P_t|\le\varepsilon$ do not trigger a change in ramp direction. Each interval must additionally satisfy a minimum cumulative ramp-magnitude threshold:

\begin{equation}
M_i := \left|P_{t_i^e}-P_{t_i^s}\right| \ge R^{th},
\label{eq:ramp_threshold}
\end{equation}
\noindent where $M_i$ denotes the total magnitude of the $i$-th ramp event; $t_i^s$ and $t_i^e$ denote its start and end times, respectively; and $R^{th}$ denotes the minimum ramp-magnitude threshold. All ramp-event evaluations in this study use $R^{th}=20\%$ of installed PV capacity, consistent with prior solar-ramp studies that identify significant solar-power ramps using a 20\%-of-capacity magnitude criterion \cite{cuiCharacterizingAnalyzingRamping2017}.

Predicted and observed ramp events are matched using a direction-constrained temporal criterion. Let the $i$-th observed event and the $j$-th predicted event be denoted by
$E_i=(t_i^s,t_i^e,d_i,M_i)$ and $\hat E_j=(\hat t_j^s,\hat t_j^e,\hat d_j,\hat M_j)$, respectively, where $d_i,\hat d_j\in\{\mathrm{up},\mathrm{down}\}$ denote ramp direction. A predicted event is considered a candidate match for an observed event only when the two events have the same direction and their temporal intervals overlap or are separated by no more than one minute. Specifically, the temporal gap is defined as
\begin{equation}
g_{ij}:=
\max\left\{
0,\,
\max\left(t_i^s,\hat t_j^s\right)
-
\min\left(t_i^e,\hat t_j^e\right)
\right\},
\label{eq:event_gap}
\end{equation}
and $(E_i,\hat E_j)$ is an admissible candidate pair if $d_i=\hat d_j$ and $g_{ij}\leq 1$ min. The one-minute tolerance corresponds to the temporal resolution of the dataset and forecast outputs.

When multiple admissible candidate pairs exist, one-to-one matching is performed by minimizing the total event-boundary discrepancy, with the pairwise matching cost defined as
\begin{equation}
C_{ij}:=
\left|\hat t_j^s-t_i^s\right|
+
\left|\hat t_j^e-t_i^e\right|.
\label{eq:event_matching_cost}
\end{equation}
Each observed or predicted event can therefore be matched at most once. Matched pairs are counted as hits, unmatched observed events as misses, and unmatched predicted events as false alarms. Ramp magnitude is not used in the matching procedure and is evaluated independently through MRME, thereby avoiding selection of matched events based on magnitude accuracy.

Using the matched events, the Critical Success Index (CSI) is calculated as

\begin{equation}
\mathrm{CSI}:=\frac{N^{hit}}{N^{hit}+N^{miss}+N^{false}},
\label{eq:csi}
\end{equation}
\noindent where $N^{hit}$, $N^{miss}$, and $N^{false}$ denote the numbers of correctly forecasted, missed, and falsely predicted ramp events, respectively. A higher CSI indicates greater event-detection accuracy.

For the $N^{hit}$ matched ramp events, Mean Start Time Error (MSTE), Mean End Time Error (METE), and Mean Relative Magnitude Error (MRME) are defined as

\begin{equation}
\mathrm{MSTE} := \frac{1}{N^{hit}}\sum_{i=1}^{N^{hit}}\left|\hat t_i^s-t_i^s\right|,\quad
\mathrm{METE} := \frac{1}{N^{hit}}\sum_{i=1}^{N^{hit}}\left|\hat t_i^e-t_i^e\right|,
\label{eq:mste_mete}
\end{equation}

\begin{equation}
\mathrm{MRME} := \frac{1}{N^{hit}}\sum_{i=1}^{N^{hit}}\frac{\left|\hat M_i-M_i\right|}{M_i},
\label{eq:mrme}
\end{equation}
\noindent where $\hat t_i^s$, $\hat t_i^e$, and $\hat M_i$ denote the predicted start time, end time, and ramp magnitude, respectively.

To quantify early-warning capability, each observed ramp is also assigned an onset lead time
\begin{equation}
L_i := t_i^s-t_0,
\label{eq:ramp_lead_time}
\end{equation}
\noindent where $t_0$ is the forecast issuance time. Observed events are stratified into four onset-lead bins, 1--4, 5--8, 9--12, and 13--16 minutes according to their true onset times. For each bin, matched events and unmatched observed events contribute to $N^{hit}$ and $N^{miss}$, respectively. An unmatched predicted event is assigned to a bin according to its predicted onset lead time $\hat L_j=\hat t_j^s-t_0$ and contributes to $N^{false}$ in that bin. MSTE, METE, and MRME are computed from the matched pairs whose observed events belong to the corresponding onset-lead bin. This event-centered stratification isolates degradation with increasing advance-warning time more directly than simply truncating the forecast horizon, which would change the population of events included in each comparison.

\subsection{Forecast-Informed Reserve Evaluation}
To quantify an operational implication of improved PV forecasts, we evaluate a forecast-informed fast upward balancing-reserve margin based on residual PV forecast errors. For forecasting method $m$, the maximum upward balancing requirement over the $H=16$ min horizon is
\begin{equation}
E_{t}^{\mathrm{up},m}:=\max_{1\le h\le H}\left[\hat P_{t+h}^{m}-P_{t+h}\right]_{+},
\label{eq:up_residual}
\end{equation}
where $[x]_{+}=\max(x,0)$. For a target reliability level $\rho$, the reserve margin is calibrated only on the validation set $\mathcal V$ as
\begin{equation}
R_{m,\rho}^{\mathrm{up}}:=Q_{\rho}\!\left(\left\{E_{t}^{\mathrm{up},m}:t\in\mathcal V\right\}\right),
\label{eq:reserve_quantile}
\end{equation}
where $Q_{\rho}(\cdot)$ denotes the empirical $\rho$-quantile. The calibrated margin is then frozen and applied to the test set, where realized reserve coverage is the fraction of forecast windows satisfying $E_{t}^{\mathrm{up},m}\le R_{m,\rho}^{\mathrm{up}}$. We evaluate $\rho\in\{0.90,0.95,0.975,0.99\}$ and report reserve requirements normalized by installed PV capacity. This metric quantifies a forecast-error reserve proxy on the evaluated subset. It does not establish system-wide reserve procurement or dispatch-cost savings. We focus on upward reserve because PV overprediction creates a generation shortfall that must be supplied by fast balancing resources.

For the stochastic VideoGPT and SkyGPT baselines, 10 future-sky samples are generated for each forecast window and passed through the corresponding PV forecasting pipeline. All downstream PV, ramp, and reserve evaluations use the pointwise median of the resulting 10 PV trajectories,
\begin{equation}
\hat P_{t+h}^{m}:=\operatorname{median}_{s=1,\ldots,10}\hat P_{t+h}^{m,(s)},
\label{eq:median_prediction}
\end{equation}
so that the deployable comparison does not rely on future ground-truth information to select an oracle sample. For the ``Best-VGG'' variants of VideoGPT and SkyGPT, the same 10 stochastic future-sky samples are generated, but the sample with the highest mean VGG-based cosine similarity to the ground-truth future frames is selected retrospectively. Because this selection uses future ground truth, Best-VGG is reported only as a non-deployable oracle reference. No probabilistic reserve model is used in this study.

\subsection{Supporting Metrics for Future Sky-Video Prediction}
Video-quality metrics are used as secondary diagnostics to assess whether the future-sky representation preserves information that is relevant to downstream PV forecasting. As summarized in TABLE~\ref{tab:video_prediction_metrics}, we report four frame-level metrics---Peak Signal-to-Noise Ratio (PSNR), Structural Similarity Index (SSIM) \cite{wang2004ssim}, Learned Perceptual Image Patch Similarity (LPIPS) \cite{zhang2018unreasonable}, and VGG-based Cosine Similarity (VGGCS) \cite{simonyan2015very}---together with Temporal Optical Flow Consistency (TOF) \cite{teed2020raft} and Temporal Feature Change Distance (TFCD). These metrics respectively characterize pixel-level prediction accuracy, structural and perceptual similarity, semantic consistency, and temporal evolution of the predicted sky sequence. Their detailed mathematical definitions are provided in the supplementary file \cite{WangYou_PVRamp_2026}.

\begin{table}[!t]
\caption{Metrics Used for Video Prediction Assessment\label{tab:video_prediction_metrics}}
\centering
\resizebox{\columnwidth}{!}{%
\begin{tabular}{@{}l l p{3.6cm} c@{}}
\hline\hline
\textbf{Metric} & \textbf{Type} & \textbf{Measures} & \textbf{Better When} \\
\hline
PSNR & Frame-based & Pixel-wise prediction accuracy. & $\uparrow$ \\
SSIM & Frame-based & Structural similarity (luminance, contrast, structure). & $\uparrow$ \\
LPIPS & Frame-based & Perceptual similarity in deep feature space. & $\downarrow$ \\
VGGCS & Frame-based & Semantic similarity (cosine in VGG feature space). & $\uparrow$ \\
TOF & Temporal & Temporal motion consistency (optical flow difference). & $\downarrow$ \\
TFCD & Temporal & Temporal feature consistency (difference in feature dynamics between consecutive frames). & $\downarrow$ \\
\hline\hline
\end{tabular}%
}
\end{table}

\section{Results of PV Forecasting and Supporting Sky-Video Evaluation}

\subsection{PV Power Forecasting Performance}
We benchmark ultra-short-term PV forecasting against ConvLSTM \cite{ShiEtAl_ConvLSTM_NIPS2015}, PhyDNet \cite{LeGuenThome_CVPRW2020_IrradianceFisheye}, VideoGPT \cite{yanVideoGPTVideoGeneration2021}, SkyGPT \cite{nieSkyGPTProbabilisticUltrashortterm2024}, and ECLIPSE \cite{palettaECLIPSEEnvisioningCLoud2022}. Deployable evaluations use the median of the stochastic forecasts from VideoGPT and SkyGPT; ``Best-VGG'' uses the retrospective ground-truth-based VGG-similarity selection defined in Section~IV-B, while ``RealImg'' provides an upper-bound reference.

\begin{figure}[!t]
\centering
\includegraphics[width=\columnwidth]{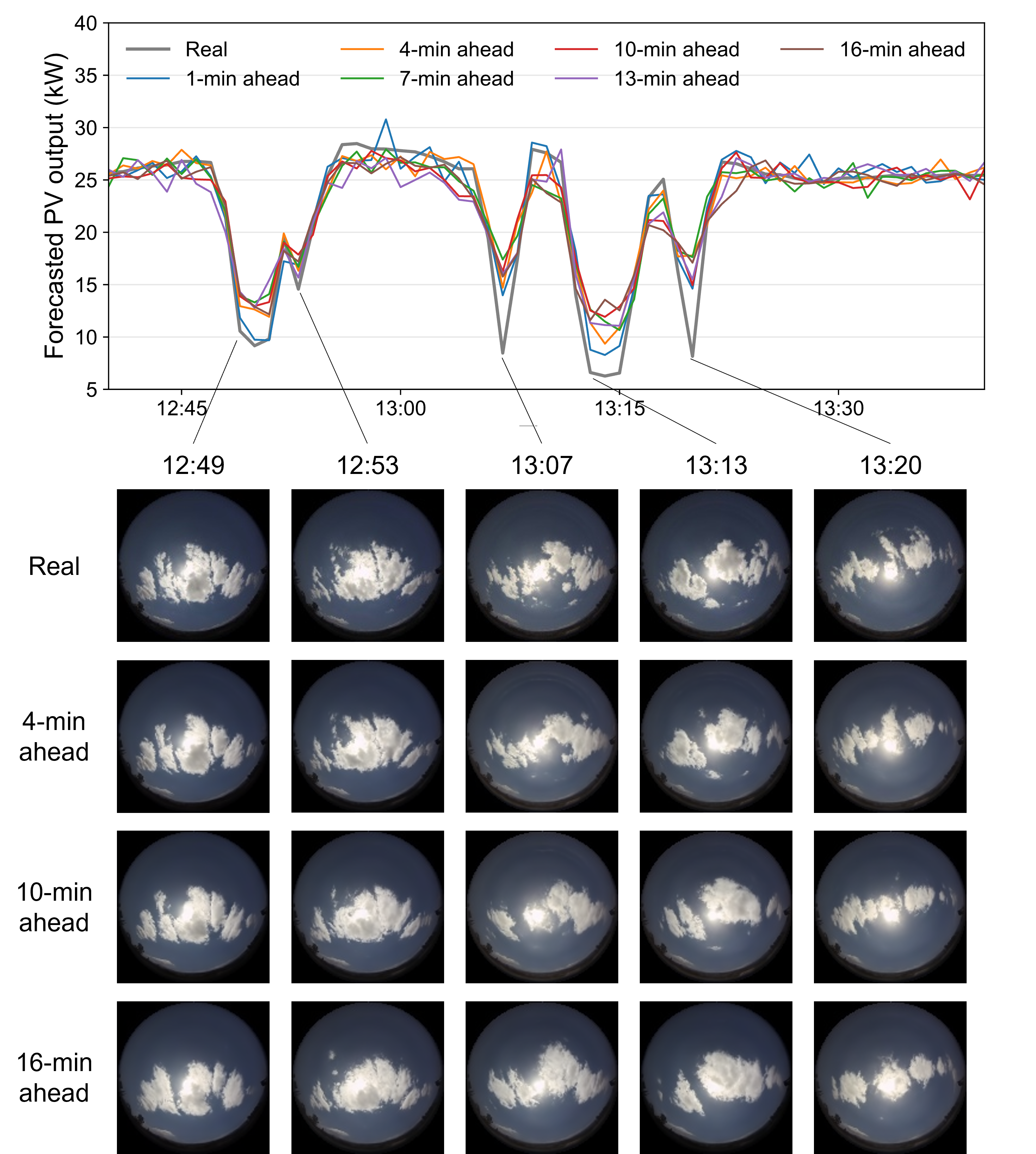}
\caption{Forecasted PV output curves using the PhyDiffNet and RaPVFormer frameworks. Predicted full-sky frames at five key ramp event time points are annotated alongside the PV curves, illustrating temporal consistency and ramp-tracking accuracy.}
\label{fig:pv_forecast_with_frames}
\end{figure}

Fig.~\ref{fig:pv_forecast_with_frames} illustrates a representative case in which the proposed framework reproduces the timing and shape of short-lived, cloud-driven PV ramps.

\begin{figure}[!t]
\centering
\includegraphics[width=\columnwidth]{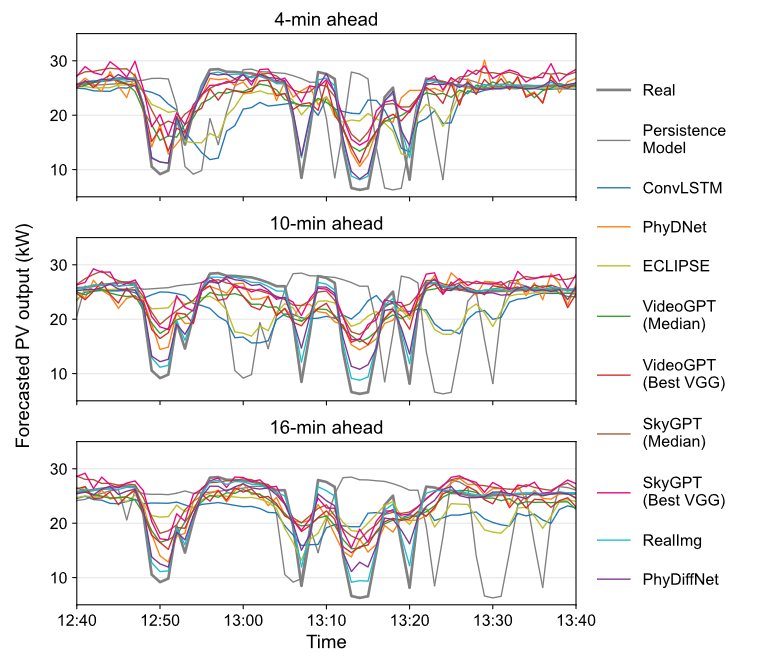}
\caption{Comparative analysis of PV output forecasts across different methods at 4-, 10-, and 16-minute lead times.}
\label{fig:comparative_forecast_methods}
\end{figure}

Fig.~\ref{fig:comparative_forecast_methods} shows that the proposed framework better preserves the timing and shape of rapid PV transitions, whereas the baselines tend to smooth or mis-time ramps.

\begin{figure}[!t]
\centering
\includegraphics[width=\columnwidth]{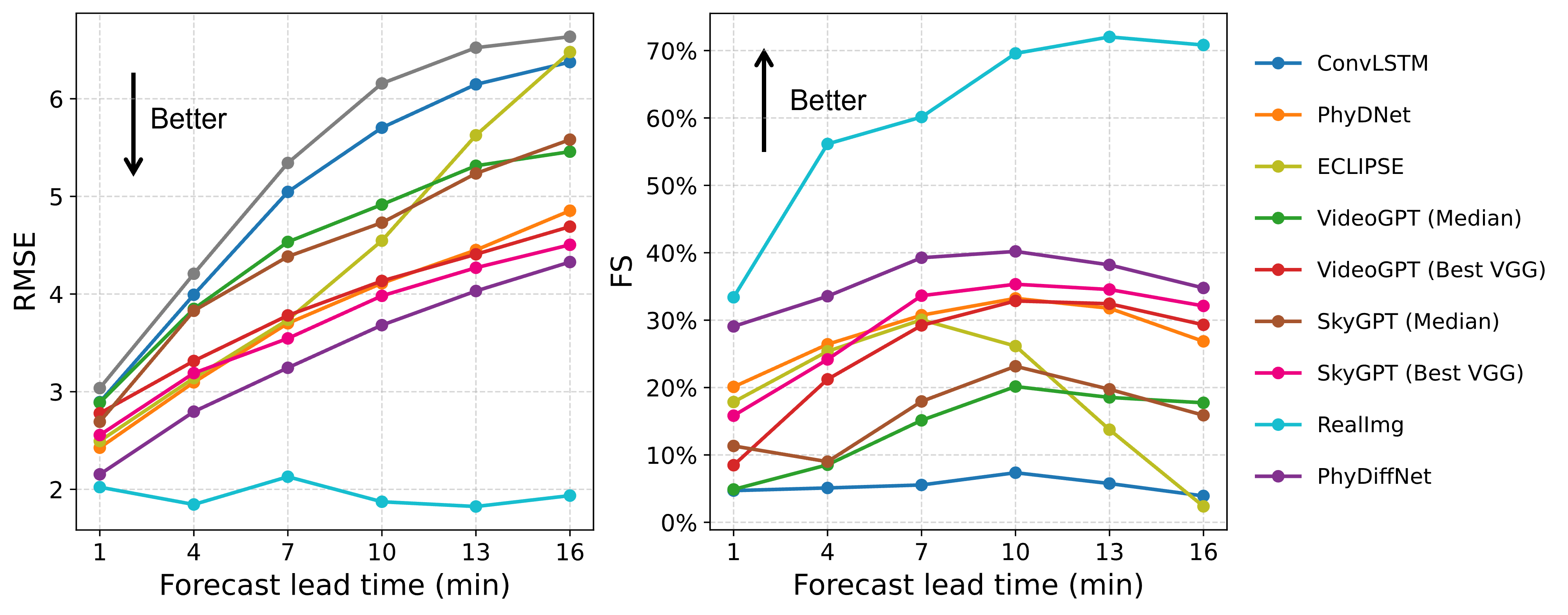}
\caption{RMSE and FS metrics for PV output forecasting across prediction horizons.}
\label{fig:rmse_fs_metrics}
\end{figure}

Fig.~\ref{fig:rmse_fs_metrics} confirms the strongest deployable PV-forecasting performance for PhyDiffNet. Forecast accuracy decreases with lead time, while the gap to RealImg indicates further potential for improving PV forecasts through more accurate future-sky prediction.

\subsection{Lead-Time-Dependent PV Ramp Event Forecasting}

Ramp-event performance is evaluated across the four onset-lead bins defined in Section~IV-A using the fixed 20\% ramp threshold \cite{cuiCharacterizingAnalyzingRamping2017}. TABLE~\ref{tab:lead_time_ramp_performance} summarizes event detection, timing, and magnitude accuracy; stochastic baselines follow the deployable median-forecast protocol, while RealImg provides an upper-bound reference.

\begin{table}[!t]
\caption{Lead-Time-Dependent PV Ramp Event Forecasting Performance\label{tab:lead_time_ramp_performance}}
\centering
\tiny
\setlength{\tabcolsep}{1.8pt}

\resizebox{\columnwidth}{!}{%
\begin{tabular}{@{}lcccccccc@{}}
\hline\hline
\textbf{Method} & \multicolumn{4}{c}{\textbf{CSI (\%) $\uparrow$}} & \multicolumn{4}{c}{\textbf{MSTE (min) $\downarrow$}} \\
\cline{2-5}\cline{6-9}
 & \textbf{1--4} & \textbf{5--8} & \textbf{9--12} & \textbf{13--16} & \textbf{1--4} & \textbf{5--8} & \textbf{9--12} & \textbf{13--16} \\
\hline
ConvLSTM & 50.7 & 48.1 & 44.1 & 39.9 & 2.45 & 3.38 & 3.89 & 5.21 \\
PhyDNet & 86.0 & 81.6 & 74.6 & 67.8 & 1.09 & 1.63 & 2.14 & 2.91 \\
ECLIPSE & 70.4 & 66.9 & 61.4 & 55.6 & 1.63 & 2.36 & 3.06 & 4.21 \\
VideoGPT (Median) & 83.1 & 77.7 & 72.2 & 63.1 & 1.28 & 1.74 & 2.26 & 3.12 \\
SkyGPT (Median) & 84.7 & 79.7 & 73.6 & 66.3 & 1.20 & 1.61 & 2.06 & 2.93 \\
PhyDiffNet (Our) & \textbf{96.2} & \textbf{93.8} & \textbf{91.5} & \textbf{84.9} & \textbf{0.35} & \textbf{0.67} & \textbf{0.93} & \textbf{1.35} \\
RealImg & 99.8 & 98.8 & 97.9 & 96.7 & 0.13 & 0.24 & 0.44 & 0.90 \\
\hline
\multicolumn{9}{@{}c@{}}{}\\[-4pt]
\textbf{Method} & \multicolumn{4}{c}{\textbf{METE (min) $\downarrow$}} & \multicolumn{4}{c}{\textbf{MRME (\%) $\downarrow$}} \\
\cline{2-5}\cline{6-9}
 & \textbf{1--4} & \textbf{5--8} & \textbf{9--12} & \textbf{13--16} & \textbf{1--4} & \textbf{5--8} & \textbf{9--12} & \textbf{13--16} \\
\hline
ConvLSTM & 2.44 & 3.28 & 4.34 & 5.55 & 30.22 & 41.11 & 52.19 & 68.17 \\
PhyDNet & 1.37 & 1.98 & 2.54 & 3.50 & 18.17 & 24.51 & 31.20 & 40.87 \\
ECLIPSE & 1.68 & 2.52 & 3.21 & 4.32 & 24.11 & 32.18 & 41.12 & 52.92 \\
VideoGPT (Median) & 1.47 & 2.00 & 2.54 & 3.44 & 22.09 & 27.31 & 34.07 & 44.94 \\
SkyGPT (Median) & 1.33 & 1.89 & 2.46 & 3.15 & 20.89 & 25.97 & 32.16 & 43.00 \\
PhyDiffNet (Our) & \textbf{0.44} & \textbf{0.66} & \textbf{0.97} & \textbf{1.59} & \textbf{9.11} & \textbf{13.41} & \textbf{18.41} & \textbf{27.05} \\
RealImg & 0.18 & 0.28 & 0.49 & 0.98 & 7.09 & 9.51 & 12.30 & 20.05 \\
\hline\hline
\end{tabular}
}
\end{table}

PhyDiffNet achieved the strongest deployable performance across all onset-lead bins, with its CSI advantage over deployable baselines becoming more pronounced at longer lead times, while the remaining gap to RealImg indicates that further improvements in future-sky prediction could extend useful ramp-warning horizons.

\subsection{Operational Implication: Reliability--Reserve Tradeoff}
\begin{figure}[!t]
\centering
\includegraphics[width=0.62\columnwidth]{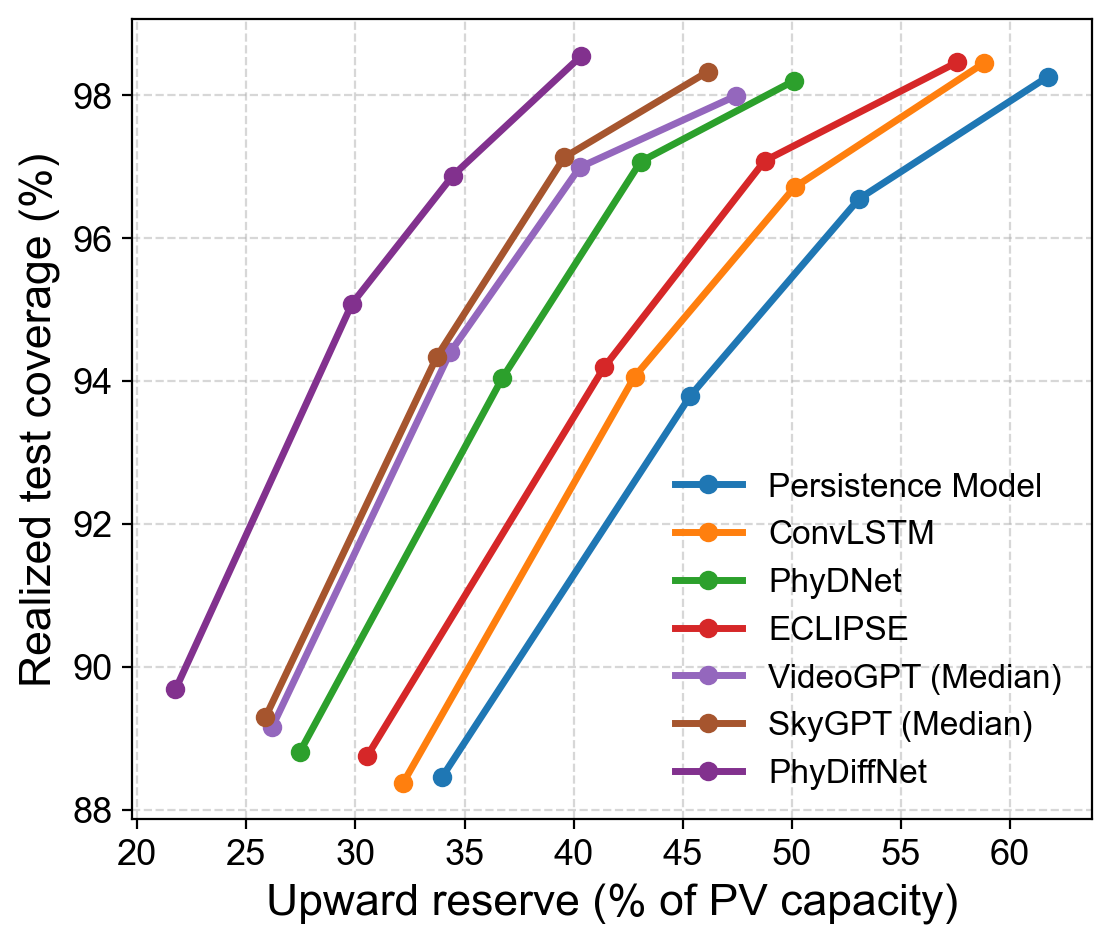}
\caption{Reliability--reserve tradeoff for forecast-informed fast upward balancing reserve. Reserve margins are calibrated on the validation set and applied unchanged to the test set. }
\label{fig:reliability_reserve_tradeoff}
\end{figure}

Fig.~\ref{fig:reliability_reserve_tradeoff} compares fast upward reserve requirements at common validation-calibrated reliability targets. The calibrated margins are applied unchanged to the held-out test set, so realized test coverage is reported separately rather than assumed to match the calibration target exactly.

At the same validation calibration target, the proposed method required a smaller upward reserve margin under the adopted residual-error metric. At the 95\% target, margins were 29.8\%, 45.3\%, and 33.7\% of rated PV capacity for PhyDiffNet, the Persistence Model, and SkyGPT (Median), respectively. SkyGPT (Median) was the strongest forecasting baseline under this reserve metric. The reductions were 34.2\% relative to the Persistence Model and 11.6\% relative to SkyGPT (Median). Realized test coverage was 95.1\%, compared with 93.8\% for persistence and 94.3\% for SkyGPT (Median). The advantage persists into the high-reliability tail, indicating that the operational implication is not confined to moderate reliability levels.

\subsection{Interpretability of PV Ramp Forecasting}

\begin{figure}[!t]
\centering
\includegraphics[width=0.6\columnwidth]{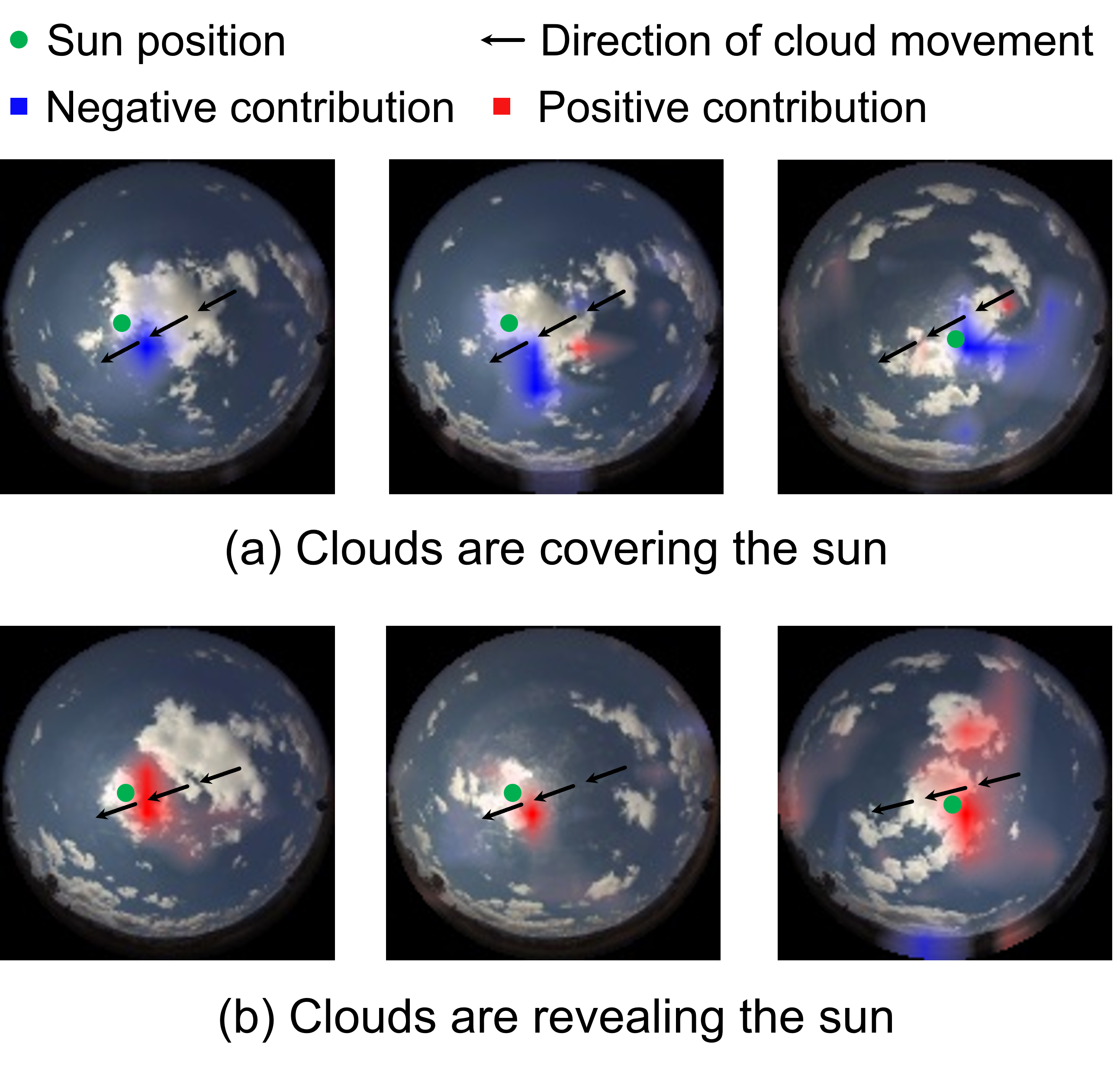}
\caption{Grad-CAM visualizations of RaPVFormer predictions, showing activation concentrated near the sun and adjacent cloud regions during occlusion and clearing events.}
\label{fig:gradcam_attention}
\end{figure}

In the illustrated cases, Grad-CAM attribution maps \cite{Selvaraju2017GradCAM} highlighted regions near the sun and neighboring clouds during occlusion and clearing (Fig.~\ref{fig:gradcam_attention}).
RaPVFormer effectively attends to cloud positions relative to the sun, highlighting key visual cues that influence solar irradiance during occlusion and clearing events. When clouds are covering the sun, the model assigns strong negative attention along cloud edges; conversely, when clouds are revealing the sun, positive attention becomes prominent.

\subsection{Supporting Evaluation of Future Sky-Video Prediction}
We next evaluate future-sky quality to examine how representation fidelity relates to downstream forecasting performance. Stochastic baselines are summarized by their median predictions, while Best-VGG is included as a retrospective, non-deployable reference.

\begin{figure}[!t]
\centering
\includegraphics[width=\columnwidth]{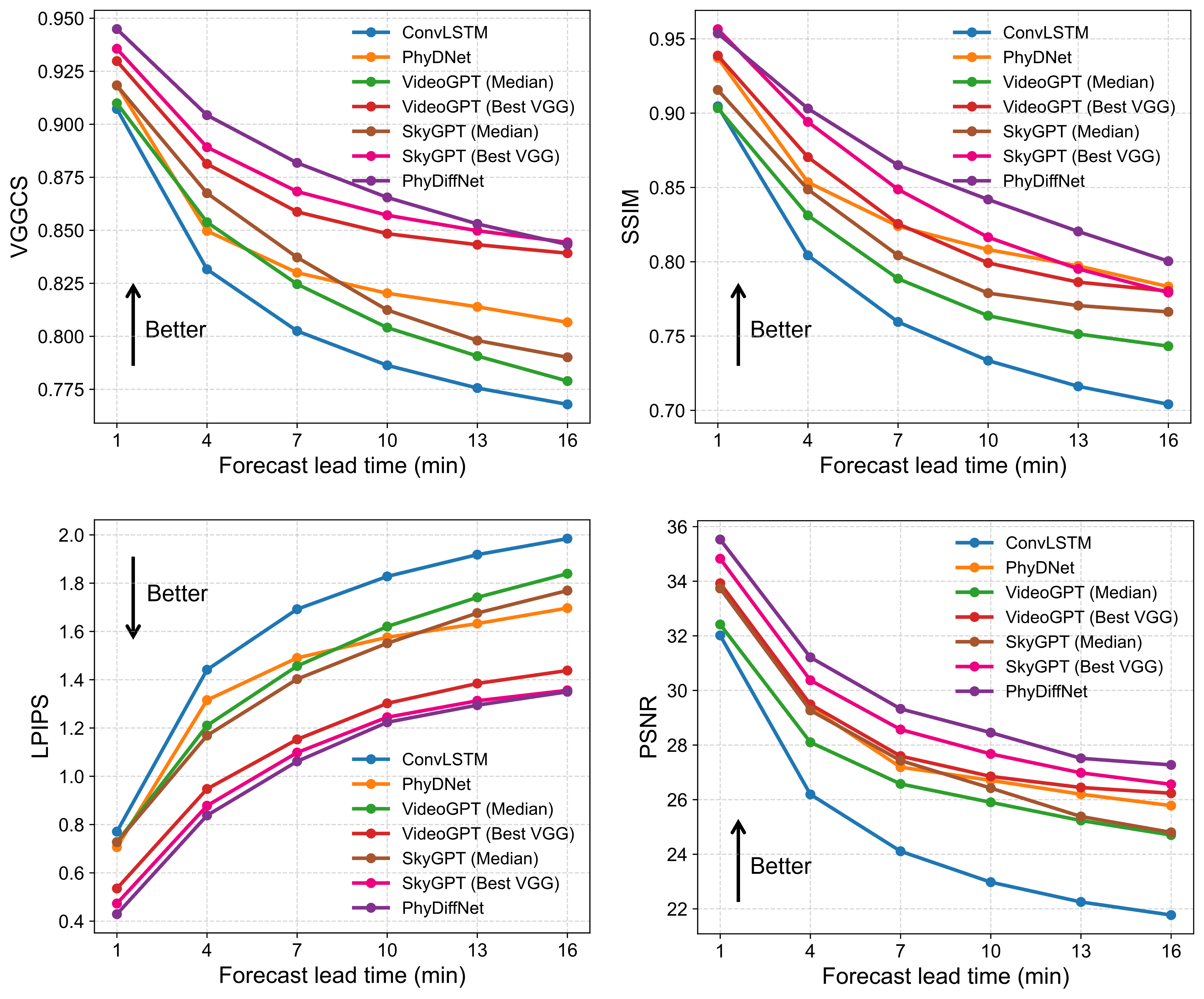}
\caption{Quantitative evaluation of predicted frames using PSNR, SSIM, LPIPS, and VGGCS metrics across time steps.}
\label{fig:frame_quality_metrics}
\end{figure}

Fig.~\ref{fig:frame_quality_metrics} shows that PhyDiffNet achieves improved frame-level reconstruction, structural, and perceptual similarity relative to the evaluated baselines.

\begin{table}[!t]
\caption{Temporal Metrics for Assessing Video Prediction Quality\label{tab:temporal_video_metrics}}
\centering
\newlength{\tablethreecolone}
\newlength{\tablethreecolother}
\setlength{\tablethreecolone}{\dimexpr0.4\columnwidth-1.6\tabcolsep\relax}
\setlength{\tablethreecolother}{\dimexpr0.3\columnwidth-1.2\tabcolsep\relax}
\begin{tabular}{@{}>{\centering\arraybackslash}m{\tablethreecolone}
                >{\centering\arraybackslash}m{\tablethreecolother}
                >{\centering\arraybackslash}m{\tablethreecolother}@{}}
\hline\hline
\textbf{Metrics} & \textbf{TOF ($\downarrow$)} & \textbf{TFCD ($\downarrow$)} \\
\hline
ConvLSTM & 1.0572 & 0.8083 \\
PhyDNet & 0.6125 & 0.7416 \\
VideoGPT (Median) & 0.7397 & 0.7502 \\
SkyGPT (Median) & 0.7272 & 0.7441 \\
PhyDiffNet (Our) & \textbf{0.3728} & \textbf{0.6276} \\
\hline\hline
\end{tabular}
\end{table}

TABLE~\ref{tab:temporal_video_metrics} shows lower TOF and TFCD values for PhyDiffNet, indicating higher temporal consistency among the evaluated methods.

\subsection{Ablation Studies}
Ablation studies isolate the contributions of the sun mask, diffusion refinement, and RaPVFormer to downstream PV forecasting. In ablation (iii), RaPVFormer is replaced with a conventional CNN-based PV forecasting model.

\begin{table}[!t]
\caption{Change of Performance in the Ablation Studies\label{tab:ablation_performance}}
\centering
\resizebox{\columnwidth}{!}{%
\begin{tabular}{@{}l c c c c c@{}}
\hline\hline
\textbf{Metrics} & \textbf{PSNR($\uparrow$)} & \textbf{SSIM($\uparrow$)} & \textbf{LPIPS($\downarrow$)} & \textbf{VGGCS($\uparrow$)} & \textbf{RMSE($\downarrow$)} \\
\hline
Original & 28.4561 & 0.8419 & 1.2237 & 0.8655 & 3.6818 \\
(i) Exclude sun-mask channel & 27.9377 & 0.8301 & 1.2541 & 0.8533 & 3.9714 \\
(ii) Exclude diffusion & 26.7077 & 0.8082 & 1.5756 & 0.8203 & 4.1108 \\
(iii) Replace RaPVFormer with CNN & 28.4561 & 0.8419 & 1.2237 & 0.8655 & 5.3762 \\
\hline\hline
\end{tabular}%
}
\end{table}

All three ablations degrade PV forecasting. Removing the sun mask and diffusion refinement increases PV RMSE from 3.6818 to 3.9714 and 4.1108, respectively, while replacing RaPVFormer with the CNN baseline increases RMSE to 5.3762. These results demonstrate that all three components contribute to the observed forecasting performance.

\section{Conclusions}
This paper shows that explicit future-cloud prediction can serve as an effective intermediate representation for event-oriented ultra-short-term PV ramp forecasting. PhyDiffNet combines physics-informed cloud-motion modeling with video-conditional diffusion, while RaPVFormer integrates observed and predicted sky states, historical PV output, and sun-location information to forecast PV trajectories and ramp events. Across increasing ramp-onset lead times, CSI decreases from 96.2\% at 1--4 minutes to 84.9\% at 13--16 minutes, demonstrating useful event-detection capability throughout the 16-minute forecasting horizon. At a validation-calibrated 95\% reliability target, the proposed forecasts reduce the forecast-error-induced fast upward reserve margin by 34.2\% relative to the Persistence Model and by 11.6\% relative to SkyGPT (Median) under the adopted reserve metric. The average end-to-end inference time is 2.62~s per forecast issuance, supporting 1-minute forecast updates.

The present evaluation focuses on the single-site SKIPP'D dataset, dynamically evolving periods that are most relevant to minute-scale PV ramps. This choice reflects the need for minute-resolution PV measurements together with unobstructed full-sky imagery that preserves detailed sun--cloud geometry; several existing public datasets do not simultaneously satisfy these requirements. Future work will extend the evaluation to additional sites and operating conditions as suitable datasets become available.


\end{document}